[1]Environmental tests of the HXI spectrometer for the ASO-S mission
Deng-yi CHEN[1, 2], Yi-ming Hu[1,*], Zhe Zhang, Yan Zhang, Yong-qiang Zhang, Jian-hua Guo, Yong-yi Huang, Tao Ma, Jin Chang

[1] Key Laboratory of Dark Matter and Space Astronomy, Purple Mountain Observatory, Chinese Academy of Sciences, Nanjing 210023, China

[2] School of Astronomy and Space Science, University of Science and Technology of China, Hefei 230026, China

Abstract: The advanced space-based solar observatory (ASO-S) is the first Chinese solar mission which is scheduled to be launched in the end of 2021. It is designed to study the magnetic field, the solar coronal mass ejections the solar flares as well as their relationships. ASO-S includes three scientific payloads, the Full-disk vector MagnetoGraph (FMG), the Lyman-alpha Solar Telescope (LST) and the Hard X-ray Imager (HXI). As a key part on ASO-S, HXI will improve our understanding on the solar flares during the 25[th] solar maximum. The HXI is composed of three instruments, the collimator (HXI-C), the spectrometer (HXI-S) and the electrical control box (HXI-E). This paper describes the procedure and results of the environmental tests including the mechanical and thermal which were done on the qualification model of the HXI-S. The functional tests of the spectrometer are further carried out, and the results show that the detector could operate normally.

Key words: ASO-S mission; HXI spectrometer; Environmental tests; Functional tests

1 Introduction

Solar flares and coronal mass ejections (CMEs) are two types of powerful activities on the sun. Huge energies would be released in short times, resulting in disastrous consequences to the space environment. It is widely believed that the energetics of both eruptions are from the magnetic field. Therefore, it is crucial to understand how the magnetic field could connect with these two phenomena of the sun. During the past 50 years, dozens of space-borne missions have been launched. However, none of these instruments could observe the solar flares, CMEs and the magnetic field simultaneously. To overcome such a problem, the advanced space-based solar observatory (ASO-S) was proposed by the Chinese solar community in 2010 and was formally applied since the autumn of 2017. The ASO-S focuses on the observation of solar flares, CMEs and the solar magnetic field at the same time during the 25[th] solar maximum (around 2025). To fulfill these science objective, it's equipped with three instruments, the Full disk vector MagnetoGraph (FMG), the Lyman Alpha Telescope (LST), and the Hard X-ray Imager (HXI). More details about the science, instrument design and software can be referred to Ref. [1-4].

As one of the key instrument on the ASO-S, the HXI adopts the similar principle as the Reuven Ramaty high-energy solar spectroscopic imager (RHESSI) and the hard X-ray telescope on Solar-A mission (HXT) for imaging [5-6]. It is a spatial modulation X-ray telescope aimed at imaging the full solar disk in hard X-rays between 30keV and 200keV. The HXI deploys three subsystems, the collimator (HXI-C), the spectrometer (HXI-S) and the electrical control box (HXI-E). The HXI-C dedicates to modulating the incident X-rays with the sub-collimators. The HXI-S is to record the photon count and measure the spectrum of these X-rays. The HXI-E is for data acquiring and

---

preliminary processing. Details of each subsystem will be described in later section 2.1.

To demonstrate the adaptability and performance of the instrument during the launch and the on-orbit operation, a series of tests should be implemented. This article concentrates on the procedures and results of the environmental tests on the HXI-S done from Dec.2019 to Mar.2020. Mechanical and thermal tests are carried out on the engineering qualification model (EQM hereafter). Functional tests with the flight hardware and software are also performed. Results obtained from the tests are very important for the final optimization of the instrument design.

2 Instrument description

2.1 Introduction of the HXI

As is stated above, HXI is made up of three sub-detectors shown in Fig.1. The HXI-C is responsible for the hard X-ray modulation with the collimator. There are 91 pairs of grids (44 pairs of sin-cos sub collimators plus a set of three sub collimators) with different pitches and position angles on the front and rear plates. They are used to supply enough u-v Fourier components for the reconstruction of images. To accurately locate the sun's center, HXI-C develops a solar aspect system. Displacement and distortion of the HXI-C are monitored by the deformation and distortion aspect system.

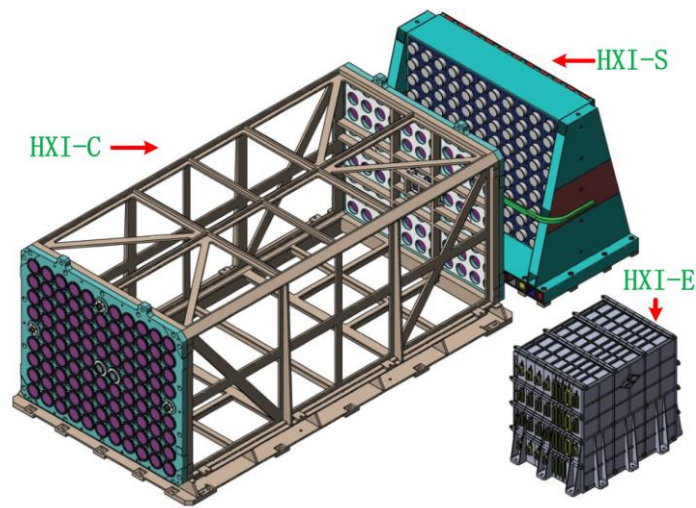

Fig.1 Schematic plot of the HXI system

The HXI-S is designed to record the photon count and measure the spectrum of the incident modulated X-rays. There are 99 detector modules in total. Besides the 91 recorders of hard X-rays, there 7 to measure the background in space 1 to charge the total flux with a thin (0.6mm) aluminum window in front. Photomultiplier tubes (PMTs) are mounted with the scintillator with a base board. The front end electronics (FEE) and high voltage distributor (HV, used for the PMT) are developed and assembled on the Faraday plate. To protect the detectors and electronic boards from potential damages during the launch and the on-orbit operation, a carbon fiber reinforced support is designed. Unlike the HXI-C and HXI-S mounted on the optical satellite plate, the HXI-E locates inside the spacecraft. It manages the system's operation, power supply, data collection, and the preliminary data processing.

Additional information about HXI-C and HXI-E can be found in Refs. [7-8];

2.2 HXI-S description

The HXI-S is an X-ray spectrometer that can absorb the energy of X-rays from 30 to 200 keV. As exploded in fig.2, it is comprised of 99 detector units arranged in an 11 ×9 array. These modules are assembled in a carbon fiber reinforced plastic structure (CFRP) which is also the interface to the satellite optical panel. X-rays recorded by the detector modules will be converted and finally transferred to the FEEs.

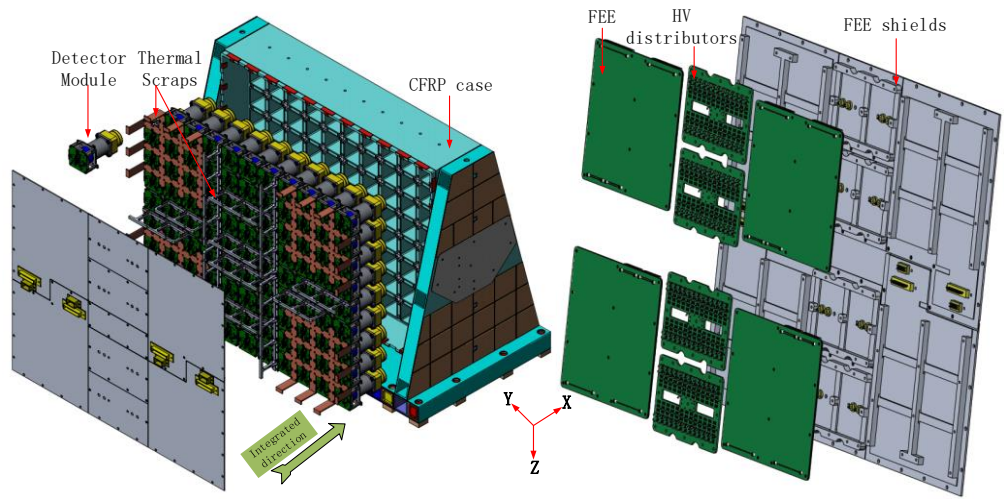

Fig. 2 Exploded view of the HXI-S

2.2.1 Detector units

The sensitive unit of the spectrometer is a Lanthanum Bromide (LaBr3 hereafter) monitored by a PMT. The LaBr3 scintillator can provide very fast light output, short decay time and an excellent energy resolution, and is thus very suitable for the X-ray detection. The cross-sectional drawing of the detector is depicted in Fig.3. The LaBr3 scintillator used in the HXI-S is produced by Beijing Glass Research Institute. Its size is φ25 mm * 25mm (cylinder height). Wrapped with the Teflon film and optical glue, LaBr3 is sealed in the aluminum case with a 3mm thickness quartz window in the rear for the transferring. The type of PMT is R1924A-100, which has the super bi-alkali cathode to achieve fine energy resolution of X-rays at ~30 keV. It's covered with 3 layers of perm-alloy sheet (about 75μm thick in total in the radial direction, to depress the magnetic field and gain variation) and placed into the protective magnesium alloy frame filled with silicone rubber Dow-coring Sylgard 170. The LaBr3 and the PMT are coupled with Dow-corning RTV615 and integrated with 4 M3 titanium screws.

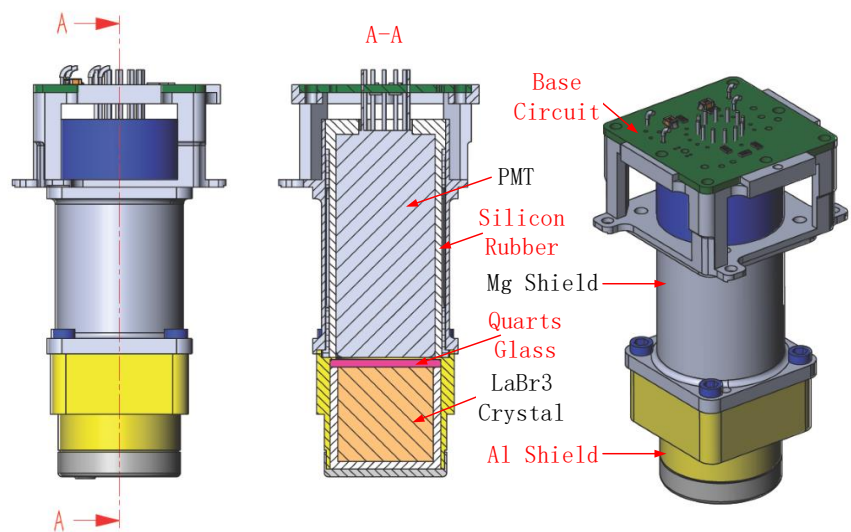

Fig.3 Profile of the HXI-S detector module

2.2.2 The CFRP case

As illustrated in Fig.3, the CFRP case is designed to support and protect the detector modules and electronics during the flight and on-orbit operation. This case should maintain them in fixed locations without degradation. At the same time the material should be as light as possible to reduce the total weight of the mission. The biggest challenge is that most of the weight budget is given to the 99 "small" modules rather than the FEEs or HVs and the support plate. Another constraint is the alignment requirement between the modules and the front sub-collimators. This requires very precise sizes and locations of the detector units without any changes.

Based on the above requirements, we adopt the CFRP (M46J with AG80 epoxy) as the main support material for the mechanical structure of the HXI-S. The CFRP provides both high strength and stiffness characteristics while its density is relatively low. From the schematic view displayed in Fig.4, the CFRP is comprised of three parts. The middle part is the key framework glued by the 99 single case, which is used to provide the location for the detectors. The sidewall supports on the left and right is designed ladder shaped and fabricated together. Closed by two carbon fiber face sheets, the sidewall provides better fixation with the middle section and stronger support capability. The bottom is another support structure composed of 20 CFRP tubes. Some inserts have been embedded inside for the purpose of not only the interface to the satellite optical panel, but also the improvement of the intensity and stiffness of the integrated structure.

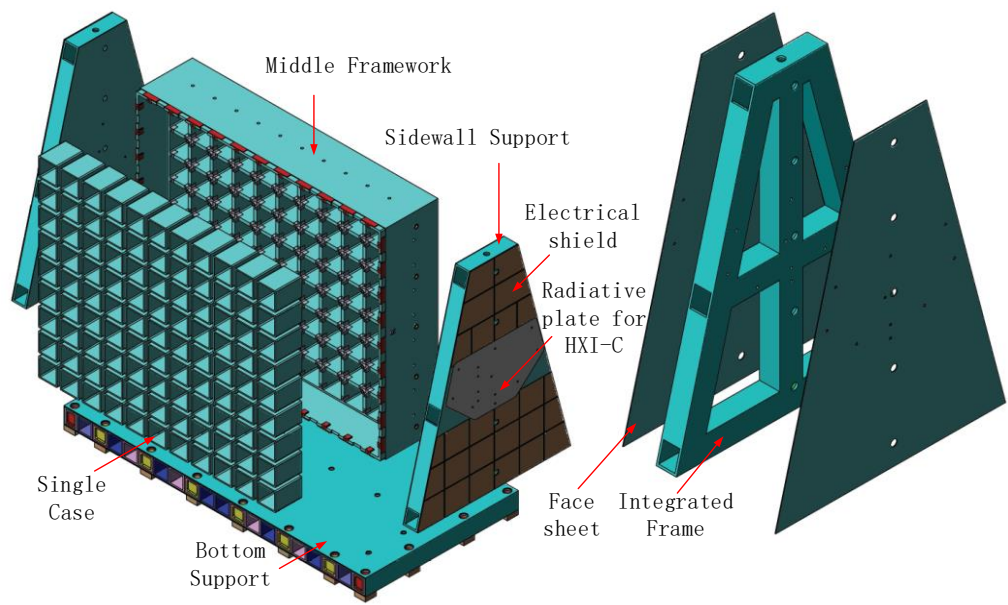

Fig.4 Schematic view of the CFRP case

Besides the detector modules and the main support, there are thermal conductive bracket located on the top of the base board (4 copper brackets and 8 aluminum), FEEs and HV distributor boards on the Faraday panel (aluminum material). To provide electrical shielding and the irradiative cold plate in space for HXI-C, an aluminum slice of 3mm thickness is glued on each sidewall.

3 Testing philosophy and procedure

Various kinds of tests are always necessary at different phases during the fabrication of the space instruments. Usually after the qualification model is developed, a series of environmental tests should be performed to verify the adaptability from the launch to the on-orbit-operation. To make the tests as realistic as possible, the HXI-S EQM uses the same drawings, materials, tooling, manufacturing processes and integration process as flight model. However, to save the cost and reduce the EQM construction time, 39 of the 99 detector modules has been replaced by the dummy model with the same mass distribution and power consumption. Thus, the EQM of the HXI-S is equivalent to a flight article in the sense of mechanical and thermal properties which are relevant for the tests.

This work presents the mechanical vibration and the thermal related tests. EMC and the remnant magnetism is not of key importance to the instrument.

3.2 Mechanical tests

Mechanical tests include the model survey, the sinusoidal and random vibration test. The model survey (by sine sweep) performs before and after each run of vibration to find the possible structural degradation induced by the test. Once the article was broken or degraded, the overall stiffness will decrease. This case could be easily figured out by the model survey. Test levels applied to the EQM of the HXI-S are summarized in Tables 1, 2 and 3.

Table 1 The sinusoidal test parameters

| Freq./Hz | 5~10 | 10~14 | 14~25 | 25~100 | 5~8 | 8~10 | 10~25 | 25~100 |
|---|---|---|---|---|---|---|---|---|
| Magnitude | 15.20mm | 6g | 9g | 3g | 23.75mm | 6g | 9g | 3g |
| Direction | X direction, 2oct/min | | | | Y direction, 2oct/min | | | |
| Freq./Hz | 5~10 | 10~40 | 40~50 | 50~75 | 75~85 | 85~100 | Z direction, 2oct/min | |
| Magnitude | 10.13mm | 4g | 8.5g | 10g | 6g | 4g | | |

Table 2 the random vibration test parameters

| Freq./Hz | 20~100 | 100~600 | 600~2000 | 9.06 RMS in each direction, |
|---|---|---|---|---|
| Magnitude | +3dB/oct | 0.1$g^2$/Hz | -9dB/oct | Duration time, 120s for each direction |

Table 3 The model survey setting

| Freq./Hz | Magnitude | direction | Sweep rate |
|---|---|---|---|
| 5～300Hz | 0.1g | X & Y | 2 oct/min |
| 5~500Hz | 0.1g | Z | |

These dynamic tests were performed at Sushi Guangbo Environmental Reliability Lab, a company specialized in space instruments vibration tests in Suzhou. The EQM was installed on the table of the shaker via a fixture in the same way mounted on the satellite optical. The fixture is rigid enough to eliminate the potential resonance with the EQM. To control the test magnitude and measure the instrument's response, several 3-axial accelerometer was positioned on the surface of the model and the fixture. The average value of the 4 monitoring-point on the fixture is adopted as the controlling strategy.

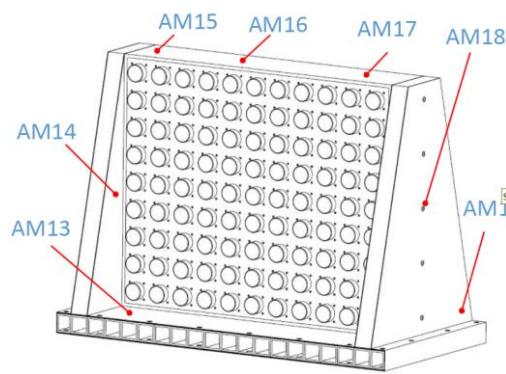
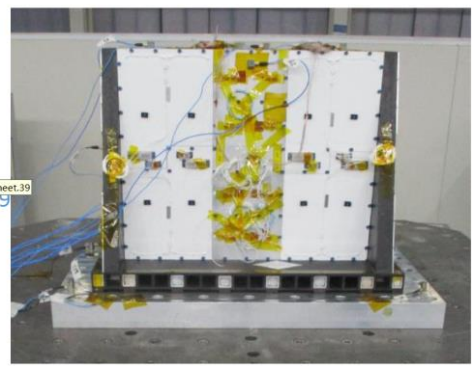

Fig.5 Distribution of some accelerometer monitors (left) and the vibration setup along Z direction (right)

Fig.6 shows an example of the response before and after the random test, measured by the accelerometers located on the CFRP case (main support) along the vertical axis. In this case, variation of the first natural frequency is less than 1% (the resonance frequency is about 255 Hz). This indicates a very good result that the support is still rigid enough without nearly no degradation.

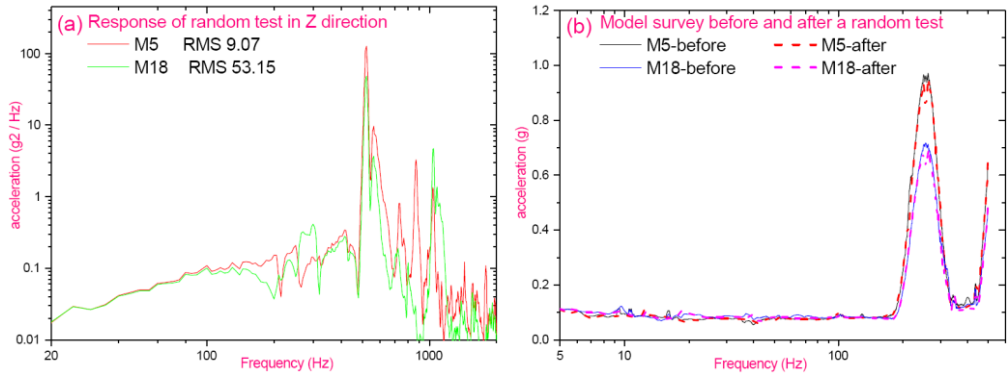

Fig.6 Response of Random vibration in the Z direction (M5 and M18)

A short functional test has also been carried out before and after the vibration test. Results will be reported in later section. Both the dynamic and fuctional verification demonstrate very reliable detector workmanship.

3.3 Thermal test

The thermal tests usually include the thermal cycles (TC), the thermal vacuum (TV) and thermal balance. The purpose of the TC is to reveal latent material defects or rosin joint in printed circuit board by loading an circulating environmental stress. It is always done at atmospheric pressure. The TV, on the other hand, is the most realistic ground simulation of the space environment. So the performance in this case would give the most direct evidence to help us judge whether the space equipment will operate regularly or not in orbit. The thermal balance, as a matter of fact, is just to validate whether the thermal design fulfills the thermal requirement or not. As for the HXI-S, only the TC and TV have been performed. The thermal balance will be carried out together with the satellite. The TC and TV test parameters are described in Table 4.

Table 4 The TC and TV test parameters

| Item | Pressure | Extreme Hot Temp. | Extreme Cold Temp. | Duration at peak | Cycling |
|---|---|---|---|---|---|
| TC | Room Pres. | +40°C | -20°C | 4h | 25.5 |
| TV | <1.3e-3 | +40°C | -20°C | 4h | 6.5 |

During the TC and TV tests, an electrical functional test was implemented all the time except the first and last loop during which the turn-on/turn-off function of the instrument will be tested either. The temperature of the spectrometer was measured by means of thermal resistors located on the surface of the detector and the inside thermal straps (4 on the copper bracket and 3 on the aluminum). Fig.8 gives an example of the temperature recorded on a detector module and the thermal straps of the TV test.

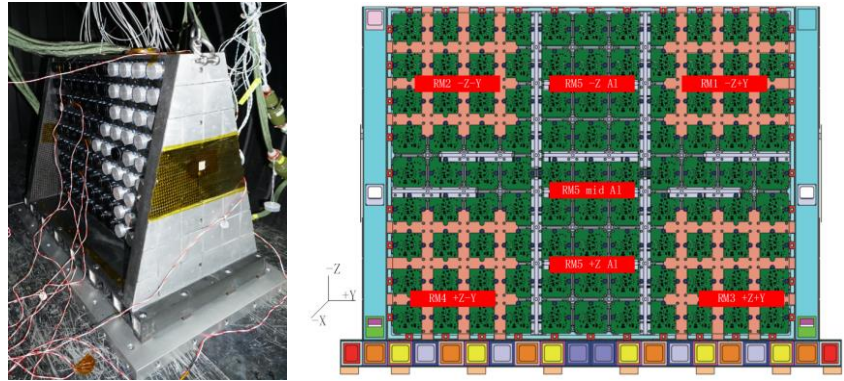

Fig.7 The TV test setup in the chamber and thermistors on the thermal scraps

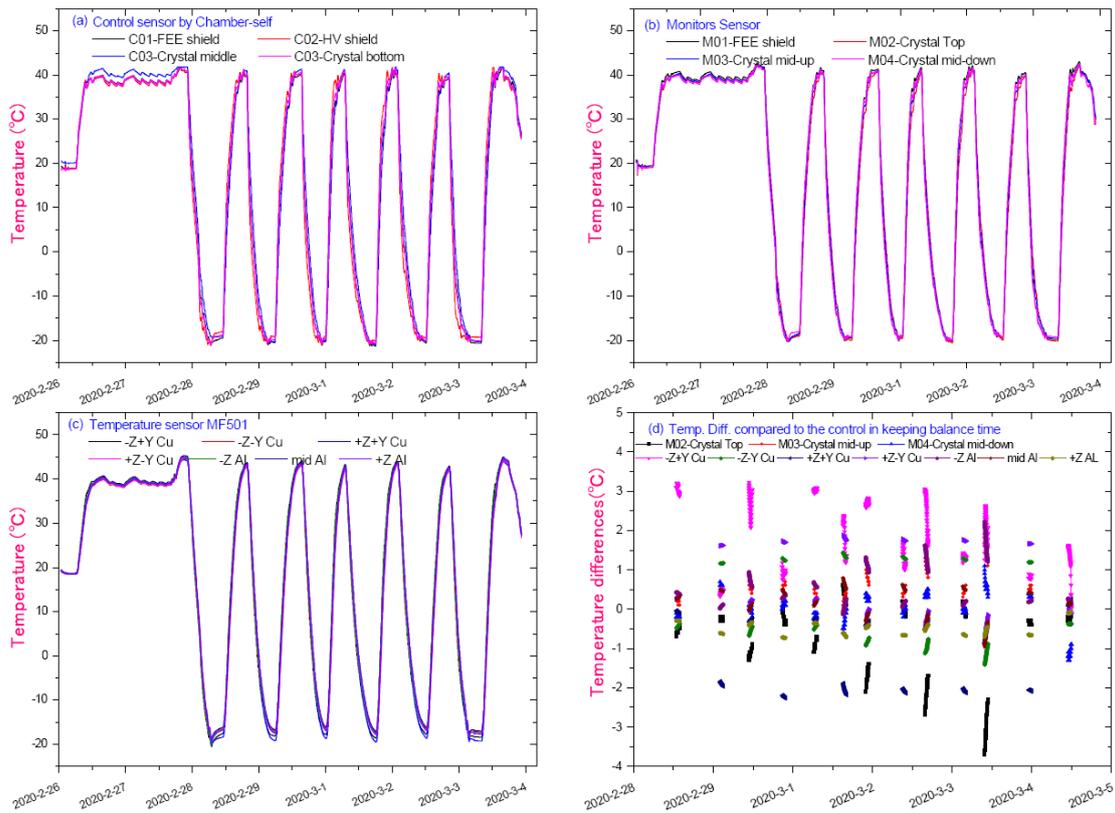

Fig.8 Temperature monitored by the sensors located on the instrument during the TV test

For the results shown in Fig.8 (a) (b) (c) panels, the start time is from the close of the chamber, and for Fig.8 (d), the start time is when the chamber vacuum pressure becomes lower than 1e-3 Pa and then last 24 hours keeping in the state. Both Fig (a), (b) and (c) imply the fine temperature control of the system. Fig.8 (d) depicts the temperature differences between the monitor points and the controlling points on the surface of the instrument. The controlling value is taken as the average value of the four points measured. Both profiles reports the whole test between -20 to +40 °C with the tolerance at ±2 °C.

The temperature distribution acquired from the test especially the TV test suggest a quite positive result. The temperature gradients are mostly less than 2 °C while very few individual points increases to about 3 °C when the monitors getting balanced. These sensors are located widely, from the detector modules to the thermal straps and to the Faraday panel (as the self-irradiative panel). Another conclusion from the test results is that the differences in the hot environment are much smaller than the cold environment. This is also good since the HXI-S will run in a temperature range from 17 °C to 27°C.

Functional test will be reported in later section 4.

4 Functional tests

Successful environmental tests on HXI-S also demands the perfect performance of the spectrometer. Thus, we test the major functions of the instrument before and after the environmental test, to make sure that the detector can operate normally in such conditions. Typically the energy resolution and the linearity dynamic range of the HXI-S are used to prove whether the instrument works well. However, LED light or charge injection from signal generator used in instrument fabrication are not possible test approach for the entire HXI-S, beam test is also impractical in environment test in environment test. The only realistic method is using the $^{133}$Ba radioactive source, which can provides more than three characteristic X-rays in 32 keV, 81keV, 356 keV and so on.

We have acquired the spectrum before and after all the test. Fig.8 presents the energy resolution of a typical detector module locates in the middle of HXI-S. In this figure spectrum peak could be found in 356keV, 81keV of X-ray from 133Ba in radioactive source, and in around 32keV with a combination of X-rays from 133Ba and 138La isotope in scintillator. The result shows that after environment test the energy resolution still maintains in 23% at 32 keV and 16% at 81 keV, almost the same before environment test.

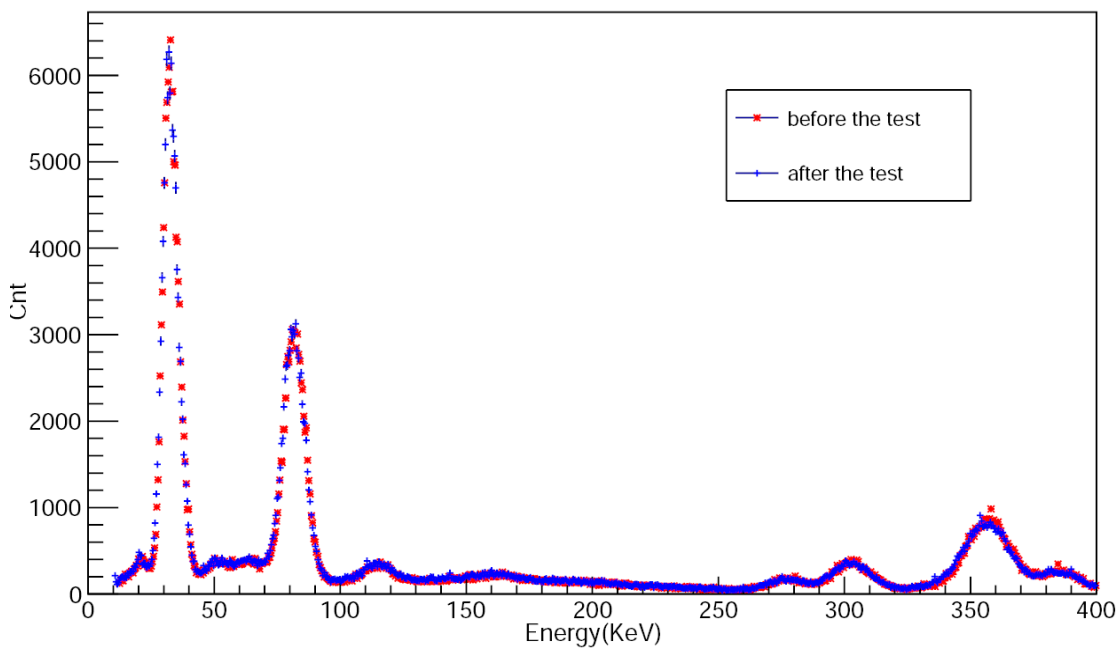

Fig.9 Energy Spectrum (Ba-133) of a typical detector module

As shown in the chart below, the linearity dynamic range is also unchanged during the environment test. The maximum integral nonlinearity (INL) is smaller than 5% with charge input range from 32keV – 356keV. It can perfectly meet the requirement of dynamic range in HXI-S.

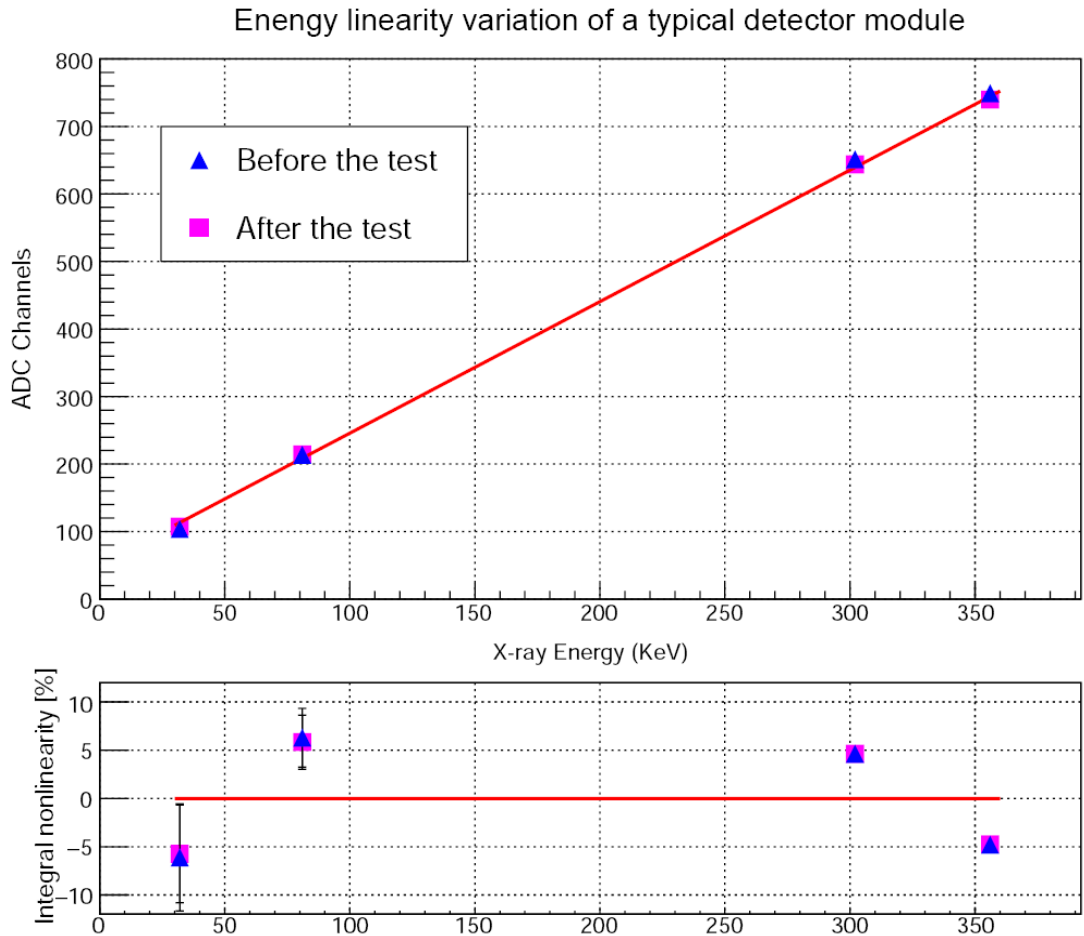

Fig.10 Energy linearity variation of a typical detector module

5 Conclusion

As the first dedicated of China, the ASO-S is undertaking the phase C (detailed design and fabrication of EQM) study and will step into phase D (flight model construction) in fall of 2020. In this sense, HXI-S has successfully gone through the mechanical and thermal tests in phase C with the EQM. In this work, we report the environmental tests done on one of the key instrument onboard the ASO-S, the HXI-S. The test results prove that the mechanical and the thermal characteristic of the ASO-S satisfies the mission requirement. The energy resolution of the spectrometer keeps good during and after the environmental tests. This study illustrates that the design and manufacture of the HXI-S instrument enables normal operation of the detector in space environment.

Acknowledgements

This work was supported by the Strategic Priority Program stage II on space science of Chinese Academy of sciences (No. XDA 15320104) and the National Natural Science Foundation of China (Nos. 11803903). We would like to thank Dr. Chunsheng Zhang and Dr. Gordon for their useful advice. The author would also appreciate the HXI team who make this work wonderful.